\documentclass[lettersize,journal]{IEEEtran}

\usepackage{ifluatex}
\usepackage{ifpdf}

\ifluatex
	\usepackage{fontspec}
	\usepackage{anyfontsize}
	\usepackage{amsmath, amsfonts, amsthm}
	
	\usepackage{unicode-math}
	
\newfontfamily\tgpfont{TeX Gyre Pagella}

\newfontfamily\arialfont{Arial}

\else
\ifpdf
	\usepackage[utf8]{inputenc}
	\usepackage{fontenc}
	\usepackage{anyfontsize}
	\usepackage{amsmath, amsfonts, amsthm}
	\usepackage{upgreek}
	\NewDocumentCommand{\symbfup}{m}{%
		\boldsymbol{\mathrm{#1}}%
	}
	\NewDocumentCommand{\symbfsfup}{m}{%
		{\bfseries\sffamily #1}%
	}
	\NewDocumentCommand{\symsfup}{m}{%
		{\textsf{#1}}%
	}
	\NewDocumentCommand{\symbb}{m}{%
		\mathbb{#1}%
	}
	\NewDocumentCommand{\symup}{m}{%
		\text{#1}%
	}
\fi
\fi
\usepackage[svgnames]{xcolor}
\usepackage[hidelinks]{hyperref}
\usepackage{thmtools}
\usepackage[nameinlink]{cleveref}
\usepackage{lipsum}
\usepackage{xparse}
\usepackage[math]{cellspace}
\usepackage{nicefrac}
\usepackage{letltxmacro}
\usepackage[breakable]{tcolorbox}
\tcbuselibrary{skins,breakable}
\tcbset{shield externalize}
\usepackage[
	backend=biber,
	style=ieee,
	doi=false,
	isbn=false,
	url=false,]{biblatex}
\usepackage{relsize}
\usepackage{tikz}
\usepackage{pgfplots}
\usepackage{caption}
\usepackage{subcaption}
\usepackage{titlesec}

\usepackage{booktabs}
\usepackage{siunitx}
\usepackage{derivative}
\usepackage{longtable}

\usepackage{algorithm}
\usepackage{algpseudocode}
\usepackage{algorithmicx}
\usepackage{array}
\usepackage{textcomp}
\usepackage{stfloats}
\usepackage{pgfplotstable}
\usepackage{hhline}
\usepackage{supertabular}

\usepackage{import}
\usepackage{ipsum}
\usepackage{orcidlink}
\usepackage{tikz-cd}
\usepackage[scr=txupr]{mathalpha}
\usepackage{tocloft}

\usepackage[normalem]{ulem}

\NewDocumentCommand{\IfEmpty}{m m G{}}{\ifthenelse{\equal{#1}{}}{#2}{#3}}%
\let\ABD\AtBeginDocument
\let\nfrac\nicefrac
\NewDocumentCommand{\MakeSymbol}{m m}{
    \ABD{\DeclareDocumentCommand{#1}{}{#2}}
}
\makeatletter
\definecolor{TodoBlue}{HTML}{50b0d0}
\definecolor{TodoGreen}{HTML}{50d060}
\definecolor{TodoOrange}{HTML}{d09030}
\DeclareDocumentCommand{\@todo}{m O{TodoOrange}}{%
	\noindent{\small\texttt{\textcolor{#2}{-TODO-:} [\\
	\phantom{aaaa}\begin{minipage}[t]{0.8\linewidth}\noindent{\color{#2}#1}\end{minipage}\\
	]}}\\%
}
\DeclareDocumentCommand{\todotext}{m}{%
	\@todo{#1}[TodoBlue]
}
\DeclareDocumentCommand{\todosym}{m}{%
	\@todo{#1}[TodoGreen]
}
\let\todo\@todo
\makeatother

\ABD{%
	\def\checkmark{
    \tikz\fill[scale=0.4](0,.35) -- (.25,0) -- (1,.7) -- (.25,.15) -- cycle;} 
	\NewDocumentCommand{\done}{s m}{%
		\IfBooleanTF{#1}{
			{\color{TodoBlue!30!white} \sout{#2}}%
		}{%
			{\color{TodoGreen!30!white}\checkmark \sout{#2}}%
		}%
	}
}

\DeclareDocumentCommand{\re}{}{%
    \symbb{R}%
}
\DeclareDocumentCommand{\im}{}{%
    \symbb{I}%
}

\AtBeginDocument{%
\DeclareDocumentCommand{\Re}{s m}{%
	\IfBooleanTF{#1}{%
		{#2}^{\re}%
	}{%
		#2^{\re}%
	}%
}
\DeclareDocumentCommand{\Im}{s m}{%
	\IfBooleanTF{#1}{%
		{#2}^{\im}%
	}{%
		#2^{\im}%
	}%
}
}

\DeclareMathOperator*{\argmin}{argmin}

\DeclareDocumentCommand{\expec}{o m}{%
	\IfValueTF{#1}{\underset{#1}{\mathbb{E}}}{\mathbb{E}}\left\{#2\right\}
}

\DeclareDerivative{\npdv}{\partial}[style-var=multiple, style-var-/=multiple, style-var-!=mixed, style-var-/!=multiple, delims-eval=(), delims-eval-/=(), delims-eval-!=(),style-frac=\nfrac]

	\MakeSymbol{\j}{\text{j}}

\MakeSymbol{\-}{
    \text{-}
}
\MakeSymbol{\dg}{^{\symsfup{o}}}
\MakeSymbol{\eqc}{\text{,}}
\MakeSymbol{\eqp}{\text{.}}
\MakeSymbol{\eqsc}{\text{;}}
\MakeSymbol{\dd}{\textrm{d}}

\MakeSymbol{\defas}{%
    \mathbin{\oset{\raisebox{1pt}{\scalebox{0.6}[0.6]{$\triangle$}}}{=}}%
}
\MakeSymbol{\div}{%
    \mathbin{//}%
}
\MakeSymbol{\mod}{%
    \mathbin{\%}%
}
\MakeSymbol{\kp}{%
    \mathbin{\otimes}%
}
\MakeSymbol{\conv}{%
    \mathbin{\ast}%
}
\MakeSymbol{\bconv}{%
	\mathbin{\circledast}%
}
\MakeSymbol{\vbconv}{%
	\mathbin{\rharpoon{\circledast}}%
}
\MakeSymbol{\e}{%
	\mathrm{e}%
}
\MakeSymbol{\naturals}{%
	\mathbb{N}%
}

\MakeSymbol{\PM}{
	\mathbin{\ooalign{\raisebox{1.5pt}{\scalebox{0.8}{$+$}}\cr\hfil\raisebox{-1.5pt}{\scalebox{0.8}{$-$}}\hfil}}
}

\MakeSymbol{\bsquare}{\fcolorbox{black}{black}{\null}}

\DeclareDocumentCommand{\review}{m g}{
    \item \IfValueTF{#2}{
         {\color{LightGray}\soutit{#1}} - #2
    }{
        {\color{Gray}\textit{#1}}
    }
}

\newcounter{reviewer}
\DeclareDocumentEnvironment{reviews}{}{
    \begin{enumerate}[label=\thereviewer.{\alph*}]
    \footnotesize
}{
    \end{enumerate}
}
\definecolor{ColA}{HTML}{991F3D}
\definecolor{ColB}{HTML}{997A1F}
\definecolor{ColC}{HTML}{3D991F}
\definecolor{ColD}{HTML}{1F997A}
\definecolor{ColE}{HTML}{1F3D99}
\definecolor{ColF}{HTML}{7A1F99}
\crefname{equation}{Eq.}{Eqs.}
\crefname{figure}{Fig.}{Figs.}
\crefname{table}{Table}{Tables}
\crefname{algorithm}{Algorithm}{Algorithms}
\crefname{section}{Section}{Sections}
\crefname{paragraph}{Paragraph}{Paragraph}
\creflabelformat{paragraph}{(#2#1#3)}

\makeatletter
\ABD{
	\DeclareDocumentCommand{\mref}{m g}{%
		\IfValueT{#2}{\cite{#2}-}\cref{#1}
	}%
}%
\ABD{
}%
\makeatother

\DeclareDocumentCommand{\extlabel}{m m O{0}m}{%
	\newlabel{#1}{{20}{1}{}{#2.#3.#4}{}}
	\newlabel{#1@cref}{{[#2][#4][]#4}{[1][1][]1}}
}

\usepgfplotslibrary{polar}
\newlength{\Yaxisshift}
\usepgfplotslibrary{colorbrewer}
\NewDocumentEnvironment{heatmap}{m O{}}{
	\begin{axis}[
		colormap/viridis,
		colorbar horizontal,
		colorbar style={
			colormap/viridis,
			width=0.6\linewidth,
			xlabel = Beampattern,
			xtick = {-10, -20, ..., -30},
			xticklabel style={
				/pgf/number format/.cd,
				fixed,
				precision=0,
				fixed zerofill,
			},
			major tick length=4pt,
			xtick style={black},
			xtick pos=top,
			extra x ticks = {0, \ymin},
			extra x tick style = {major x tick style={draw=none}},
			point meta min=\ymin,
			point meta max=0,
		},
		width=1\linewidth,
		height=0.7\linewidth,
		point meta min=\ymin,
		point meta max=0,
		mesh/cols=#1,
		xlabel style = {anchor=north, yshift=0.8\linewidth-2.0em},
		view={0}{90},
		xtick = {-180, -90, ..., 180},
		xtick style = {black},
		xticklabel style = {yshift=-3pt},
		xticklabel={$\pgfmathprintnumber{\tick}\dg$},
		grid style = {draw=none},
		ylabel style={yshift=-\linewidth+0.2em},
		ytick = {0, 8, ..., 32},
		major tick length=2pt,
		ytick style = {black},
		ytick pos = left,
		mesh/ordering=x varies,
		#2
		]
		\centering
	}{
	\end{axis}
}

\DeclareDocumentCommand{\heatmapfig}{O{0.8\linewidth} O{#7} O{\meshcols} O{\meshrows} m m m O{}}{
	\begin{subfigure}{#1}
		\centering
		\begin{tikzpicture}
			\begin{heatmap}{\meshcols}[colorbar to name={#2}, #8]
				\addplot3[surf, mesh/cols=#3, mesh/rows=#4, shader=interp] table[x=ang, y=freq, z=val, col sep=comma] {#5};
			\end{heatmap}
		\end{tikzpicture}
		\vspace*{-2mm}\caption{#6}
		\label{subfig:#7}
		\vspace*{2mm}
	\end{subfigure}
}

\newlength{\wdtwenty}
\pgfplotsset{
	every colorbar global/.append style={
		zmin=,zmax=,
	}
}

\NewDocumentEnvironment{lineplot}{O{axis} m m O{}}{ 
	\begin{#1}[
		width=1.05\linewidth,
		height=1.1\linewidth,
		xtick pos=bottom,
		ytick pos=left,
		xlabel = #2,
		ylabel = #3,
		yticklabel style={text width=3em, align=right},
		ylabel style={yshift=-2em},
		legend cell align={left},
		legend style={
			fill=white,
			draw opacity=1,
			text opacity=1,
			legend columns=4,
			/tikz/every even column/.append style={column sep=1em}
		},
		#4,
		]
		\draw[thin, dash pattern = {on 4pt off 1pt}] (axis cs:\pgfkeysvalueof{/pgfplots/xmin},0) -- (axis cs:\pgfkeysvalueof{/pgfplots/xmax},0);
}{
	\end{#1}
}

\tikzstyle{axis} = [
black
]
\tikzstyle{styleA} = [
	ColA!90!black, 
	dash pattern = {on 5pt off 0pt},
	very thick,
	mark=+, 
	mark size=2, 
	mark options={
		rotate=45,
		solid,
		very thick,
	}
]
\tikzstyle{styleB} = [
	ColB!90!black, 
	dash pattern = {on 1.2pt off 0.6pt},
	very thick,
	mark=triangle*,
	mark size=1.5, 
	mark options={
		rotate=0,
		solid,
	}
]
\tikzstyle{styleC} = [
	ColC!90!black, 
	dash pattern = {on 4pt off 1.2pt},
	very thick,
	mark=star, 
	mark size=2,
	mark options={
		rotate=0,
		solid,
	}
]
\tikzstyle{styleD} = [
	ColD!90!black,
	dash pattern = {on 4pt off 2.4pt on 1.2pt off 2.4pt},
	very thick,
	mark=square*, 
	mark size=1, 
	mark options={
		rotate=0,
		solid,
	}
]
\tikzstyle{styleE} = [
	ColE!90!black, 
	dash pattern = {on 4pt off 1.2pt on 1.2pt off 1.2pt},
	very thick,
	mark=diamond*, 
	mark size=1, 
	mark options={
		rotate=0,
		solid,
	}	
]

\pgfdeclareplotmark{starmark}{
    \node[star,star point ratio=2.25,minimum size=6pt,
          inner sep=0pt,draw=black,solid,fill=red] {};
}

\def\algcommsymb{$\#$ }
\algrenewcommand\algorithmiccomment[1]{ \hfill{\color{Gray} \algcommsymb \textit{#1}}}

\algdef{SE}[DOWHILE]{Do}{doWhile}{\algorithmicdo}[1]
{\algorithmicwhile\ #1}%

\DeclareSIUnit{\sidb}{%
	\text{dB}
}

\def\dB{\si{\sidb}}

\usetikzlibrary{positioning,calc}
\usetikzlibrary{decorations.pathreplacing}
\usetikzlibrary{shapes,shapes.geometric,shapes.arrows,decorations.pathmorphing}
\usetikzlibrary{matrix,chains,scopes,positioning,arrows,fit,backgrounds}
\usetikzlibrary{decorations.markings}
\usetikzlibrary{pgfplots.statistics, pgfplots.colorbrewer} 

\newlength{\minwidth}
\tikzstyle{basicblock} = [rectangle, minimum width=\minwidth, minimum height=1cm, text centered, draw=black]
\tikzstyle{input} = [basicblock, rounded corners, fill=blue!10]
\tikzstyle{output} = [basicblock, rounded corners, minimum width=2cm, fill=green!10]
\tikzstyle{minprocess} = [basicblock, fill=orange!10]
\tikzstyle{filtprocess} = [basicblock, fill=red!10]
\tikzstyle{decision} = [basicblock, chamfered rectangle, fill=yellow!80!red!10!white]

\tikzstyle{superinput} = [input, fill=blue!20]
\tikzstyle{superminprocess} = [minprocess, fill=orange!20]
\tikzstyle{superfiltprocess} = [filtprocess, fill=red!20]
\tikzstyle{superoutput} = [output, fill=green!20]

\tikzstyle{arrow} = [thick,->,>=stealth]

\tikzstyle{wavefront} = [%
    draw,thick,densely dotted,decorate,decoration={%
        markings, mark = between positions 0 and 1 step 2mm with {
        \draw[-,#1] (-0.03,0.06) -- (0,0) -- (-0.03,-0.06);
        }
    }
]
        
\tikzstyle{reflection} = [solid,thin,wavefront={#1}]
\input{setup/common/c0.my_accents.tex}

\newcommand{\pts}[1]{{\udel{#1}}}
\newcommand{\bts}[1]{\udel[{[}]{#1}[{]}]}
\newcommand{\cts}[1]{\udel[{\{}]{#1}[{\}}]}
\newcommand{\abs}[1]{\udel[|]{#1}[|]}

\newsavebox{\jcsboxA}
\newsavebox{\jcsboxB}
\makeatletter
\newlength{\udel@lenA}
\newlength{\udel@lenB}
\NewDocumentCommand{\udel}{O{(} m O{)}}{%
	\left#1 #2 \right#3
}%
\makeatother

\input{setup/common/c2.my_matrix_operators.tex}
\NewDocumentCommand{\SubSize}{m m}{\IfValueT{#1}{_{\scalebox{0.7}{$ \sz{#1\!}{\!\IfValueTF{#2}{#2}{#1}} $}}}}

\newlength{\SMtblimit}
\NewDocumentEnvironment{smatrix}{O{\SMtblimit}}{
	\if@display
		\setlength{\arraycolsep}{3pt}%
		\setlength{\cellspacetoplimit}{#1}%
		\setlength{\cellspacebottomlimit}{#1}%
	\fi
	~\begin{matrix} }{
	\end{matrix}~
}
\NewDocumentEnvironment{sbmatrix}{O{\SMtblimit}}{
	\setlength{\arraycolsep}{3pt}%
	\setlength{\cellspacetoplimit}{#1}%
	\setlength{\cellspacebottomlimit}{#1}%
	\left[~\begin{matrix} }{
	\end{matrix}~\right]
}
\NewDocumentEnvironment{scmatrix}{O{\SMtblimit}}{
	\setlength{\arraycolsep}{3pt}%
	\setlength{\cellspacetoplimit}{#1}%
	\setlength{\cellspacebottomlimit}{#1}%
	\left\{~\begin{matrix} }{
	\end{matrix}~\right\}
}
\NewDocumentCommand{\otup}{O{}m o m o o}{%
	{%
		\begin{sbmatrix}%
			#2#1 & \cdots#1 & \IfValueT{#3}{#3#1 & \cdots#1 & } #4
		\end{sbmatrix}\,%
	}\SubSize{#5}{#6}%
}%
\NewDocumentCommand{\ovtup}{m o m o o}{
	{%
		\begin{sbmatrix}%
			#1 \\
			\vdots \\
			\IfValueT{#2}{%
				#2 \\
				\vdots \\
			}%
			#3
		\end{sbmatrix}\,%
	}\SubSize{#4}{#5}%
}%
\NewDocumentCommand{\pair}{m m o o}{%
	{%
		\begin{sbmatrix}%
			#1 & #2
		\end{sbmatrix}\,%
	}\SubSize{#3}{#4}%
}%
\NewDocumentCommand{\vpair}{m m o o}{%
	{%
		\begin{sbmatrix}%
			#1 \\[0.5em]
			#2
		\end{sbmatrix}\,%
	}\SubSize{#3}{#4}%
}%
\def\setsep{}
\newcommand{\setprocessor}[1]{
	\setsep\IfEmpty{#1}{\cdots}{#1}
	\global\def\setsep{\,,&}
}
\newcommand{\vsetprocessor}[1]{
	\setsep\IfEmpty{#1}{\vdots}{#1}
	\global\def\setsep{\\}
}

\NewDocumentCommand{\tup}{ >{\SplitList{,}}m o o }{
    \global\def\setsep{}
	\begin{sbmatrix}%
		\ProcessList{#1}{\setprocessor}
	\end{sbmatrix}\SubSize{#2}{#3}
	\global\def\setsep{}
}
\NewDocumentCommand{\vtup}{ >{\SplitList{,}}m o o }{
    \global\def\setsep{}
	\begin{sbmatrix}%
		\ProcessList{#1}{\vsetprocessor}
	\end{sbmatrix}\SubSize{#2}{#3}
	\global\def\setsep{}
}

\DeclareDocumentCommand{\added}{o m}{%
    {
    \color{LimeGreen} \IfValueT{#1}{\textit{[#1]} }%
    #2}%
}
\DeclareDocumentCommand{\removed}{o m}{%
    {
    \color{Red} \IfValueT{#1}{\textit{[#1]} }%
    \sout{#2}}%
}
\DeclareDocumentCommand{\changed}{o m m}{%
    {
    \color{Orange} \IfValueT{#1}{\textit{[#1]} }\textit{\sout{#2}}%
    #3}%
}

\def\ieee{}

\def\todo{\color{Green}}

\makeatletter
\newcommand{\oset}[2]{{\mathpalette\o@set{{#1}{#2}}}}
\newcommand{\o@set}[2]{\o@@set{#1}#2}
\newcommand{\o@@set}[3]{%
  \vbox{\offinterlineskip
    \ialign{\hfil##\hfil\cr
      $\m@th\o@set@demote{#1}#2$\cr
      \noalign{\vskip0.2pt}
      $\m@th#1#3$\cr
    }%
  }%
}
\newcommand{\o@set@demote}[1]{%
  \ifx#1\displaystyle\scriptstyle\else
  \ifx#1\textstyle\scriptstyle\else
  \scriptscriptstyle\fi\fi
}
\makeatother

\MakeSymbol{\x}{\symsfup{x}}
\MakeSymbol{\y}{\symsfup{y}}
\MakeSymbol{\z}{\symsfup{z}}
\MakeSymbol{\w}{\omega}
\MakeSymbol{\an}{\symup{an}}
\MakeSymbol{\t}{\theta}
\MakeSymbol{\td}{\theta_{\mrm{d}}}
\MakeSymbol{\tB}{\theta_{\mrm{B}}}
\MakeSymbol{\Nu}{\mathcal{V}}
\MakeSymbol{\rn}{\symup{rn}}
\MakeSymbol{\+}{{}^{+}}
\MakeSymbol{\-}{{}^{-}}
\MakeSymbol{\Agemo}{\text{\rotatebox[origin=c]{180}{\ensuremath{\Omega}}}}

\MakeSymbol{\tD}{\tilde{D}}
\NewDocumentCommand{\wng}{O{}}{%
	\text{WNG}_{#1}%
}
\NewDocumentCommand{\df}{O{}}{%
	\text{DF}_{#1}%
}
\makeatletter
\NewDocumentCommand{\@OneIdx}{m m O{;} m}{
    #1_{\IfValueT{#2}{#2#3} {#4}}
}
\NewDocumentCommand{\@TwoIdx}{s m m m m}{
    #2_{\IfValueT{#3}{#3;}(#4\IfValueT{#3}{_{#3}}\IfBooleanT{#1}{'},#5\IfValueT{#3}{_{#3}}\IfBooleanT{#1}{'})}
}
\NewDocumentCommand{\@FourIdx}{m m m m m m}{
    #1_{\IfValueT{#2}{#2;}(#3\IfValueT{#2}{_{#2}},#4\IfValueT{#2}{_{#2}});(#5\IfValueT{#2}{_{#2}},#6\IfValueT{#2}{_{#2}})}
}

\NewDocumentCommand{\dx}{g}{
    \@OneIdx{\delta}{#1}{\x}
}
\NewDocumentCommand{\dy}{g}{
    \@OneIdx{\delta}{#1}{\y}
}
\NewDocumentCommand{\Mx}{g}{
    \@OneIdx{M}{#1}[,]{\x}
}
\NewDocumentCommand{\My}{g}{
    \@OneIdx{M}{#1}[,]{\y}
}

\DeclareDocumentCommand{\P}{g O{m}}{
    \@OneIdx{P}{#1}{#2}
}
\DeclareDocumentCommand{\S}{g O{m}}{
    \@OneIdx{S}{#1}{#2}
}
\DeclareDocumentCommand{\R}{g O{m}}{
    \@OneIdx{r}{#1}{#2}
} 
\DeclareDocumentCommand{\p}{g O{m}}{
    \@OneIdx{\psi}{#1}{#2}
}
\DeclareDocumentCommand{\D}{g O{m}}{
    \@OneIdx{D}{#1}{#2}
}
\DeclareDocumentCommand{\T}{g O{m}}{
    \@OneIdx{T}{#1}{#2}
}
\LetLtxMacro{\d}{\udot}
\DeclareDocumentCommand{\d}{g O{m}}{
    \@OneIdx{d}{#1}{#2}
}
\makeatother

\makeatletter
\DeclareDocumentCommand{\bv}{m O{n} g}{
	\@my{#2}{\symbfup{#1}}\IfValueT{#3}{_{#3}}
}
\DeclareDocumentCommand{\BV}{m g}{
    \symbfsfup{#1}\IfValueT{#2}{_{#2}}
}
\DeclareDocumentCommand{\newbv}{s o m}{
    \IfBooleanTF{#1}{
        \IfValueTF{#2}{
            \@namedef{bv#2}{\BV{#3}}
        }{
            \@namedef{bv#3}{\BV{#3}}
        }
    }{
        \IfValueTF{#2}{
            \@namedef{bv#2}{\bv{#3}}
        }{
            \@namedef{bv#3}{\bv{#3}}
        }
    }
}
\@namedef{:}{\scalebox{1.01}{\textbf{:}}}
\makeatother

\newbv{d}
\newbv{y}
\newbv{x}
\newbv{v}
\newbv{h}
\newbv[Ga]{\Gamma}
\def\Gzp{\bvGa{0,\pi}}
\newbv{T}
\newbv[La]{\Lambda}
\newbv{t}
\newbv{P}

\newbv{I}
\let\Id\bvI

\NewDocumentCommand{\Corr}{m}{%
    \bv{\Phi}_{#1}%
}

\NewDocumentCommand{\dsdi}{o}{%
	\nu\IfValueT{#1}{_{#1}}%
}

\DeclareMathOperator{\rsd}{RSD}
\DeclareMathOperator{\sub}{SUB}
\DeclareMathOperator{\tun}{TUN}
\DeclareMathOperator{\lkp}{KP}
\DeclareMathOperator{\ckp}{CKP}

\let\mrm\symup
\def\mtimes{\mathbin{\scalebox{0.7}{$\times$}}}
\NewDocumentCommand{\sz}{s m m}{%
	\IfBooleanTF{#1}{%
		\bts{{#2} \mtimes {#3}}%
	}{%
		{#2} \mtimes {#3}%
	}%
}

\NewDocumentCommand{\el}{m g O{m} o}{%
	\bts{#1}_{%
		\IfValueTF{#4}{%
			(#3\IfValueT{#2}{_{#2}},#4\IfValueT{#2}{_{#2}})%
		}{%
			#3\IfValueT{#2}{_{#2}}%
		}        
	}
}

\NewDocumentCommand{\FT}{m O{(\w)}}{%
	\symcal{F}\cts{#1}#2%
}
\NewDocumentCommand{\IFT}{m O{(t)}}{%
	\symcal{F}^{-1}\cts{#1}#2%
}

\NewDocumentCommand{\RFT}{m O{(\w)}}{%
	\symcal{S}\cts{#1}#2%
}
\NewDocumentCommand{\IRFT}{m O{(t)}}{%
	\symcal{S}^{-1}\cts{#1}#2%
}
\DeclareDocumentCommand{\S}{}{%
	\symcal{S}
}

\NewDocumentCommand{\STFT}{m O{[l,k]}}{%
	\symbb{F}\cts{#1}#2%
}
\NewDocumentCommand{\ISTFT}{m O{[n]}}{%
	\symbb{F}^{-1}\cts{#1}#2%
}

\NewDocumentCommand{\SSBT}{m O{[l,k]}}{%
	\symbb{S}\cts{#1}#2%
}
\NewDocumentCommand{\ISSBT}{m O{[n]}}{%
	\symbb{S}^{-1}\cts{#1}#2%
}

\NewDocumentCommand{\opt}{s}{%
	\IfBooleanTF{#1}{^\dagger}{^\star}
}
\NewDocumentCommand{\cost}{s g}{%
    \IfBooleanTF{#1}{{F}}{\mathscr{F}}\IfValueT{#2}{\pts{#2}}%
}

\NewDocumentEnvironment{subalign}{g}{%
	\subequations%
		\IfValueT{#1}{\label{#1}}%
		\allowdisplaybreaks%
		\align%
	}{%
		\endalign%
	\endsubequations%
}%
\NewDocumentEnvironment{subgather}{g}{%
	\subequations%
		\IfValueT{#1}{\label{#1}}%
		\allowdisplaybreaks%
		\gather%
	}{%
		\endgather%
	\endsubequations%
}%
\NewDocumentEnvironment{equations}{g}{%
	\equation%
		\IfValueT{#1}{\label{#1}}%
		\allowdisplaybreaks%
		\aligned%
	}{%
		\endaligned%
	\endequation%
}

\NewDocumentEnvironment{proposition}{m}{%
	\noindent\textbf{Proposition:} {\itshape #1}
	
	\noindent\textbf{Proof:}\itshape
}{

	\noindent\bsquare
}

\NewDocumentEnvironment{example}{}{%
	\noindent\textbf{Example:}
}{

	\noindent\bsquare
}

\newcounter{smethod}
\newcounter{method}
\NewDocumentEnvironment{method}{s}{%
    \IfBooleanT{#1}{\stepcounter{method}}%
    \stepcounter{smethod}%
	{\noindent\textbf{Method \Alph{method}.\arabic{smethod}:}}
 
}{
    \vspace{-1em}\newline%
	\noindent\bsquare%
 
}

\newtheoremstyle{break}
{9pt}
{9pt}
{}
{}
{\bfseries}
{.}
{ }
{}
\theoremstyle{break}

\newtheorem{thm}{Theorem}
\crefname{thm}{Theorem}{Theorems}

\def\square{
	\tikz\fill[scale=0.25](0,0) -- (1,0) -- (1,1) -- (0,1) -- cycle;}
\newcommand{\tmark}{%
	{{\unskip\nobreak\vadjust{\nobreak}\hskip0pt\penalty50 \space
			\vadjust{}\nobreak\hfil\square%
			\clubpenalty=0 \widowpenalty=0 \brokenpenalty=0
			\parfillskip=0pt \finalhyphendemerits=0 \par
			\penalty 10000 \parskip=0pt\noindent}}\ignorespaces} 

\NewDocumentEnvironment{theorem}{m O{}}{
	\begin{thm}{\itshape #1}#2
		
		}{
			\newline\tmark
	\end{thm}
}

\begin{document}

\title{Comparative analysis of WNG-DF compromising beamformers}
\NewDocumentCommand{\orcid}{m}{%
	$^{\orcidlink{#1}}$%
}

\author{%
	Vitor G. P. Curtarelli\orcid{0009-0009-3996-5452}
	\thanks{Manuscript Info. Corresponding author: Vitor G. P. Curtarelli.}%
	\thanks{V. G. P. Curtarelli is with the Electrical and Engineering Department at Universidade Federal de Santa Catarina (UFSC), Florianópolis, SC, Brazil (email: \url{vitor.curtarelli@posgrad.ufsc.br}).}%
} 

\markboth{IEEE SIGNAL PROCESSING LETTERS, Vol. X, 20YY}%
{Curtarelli et al.: SVD and NCM joint estimation}

\IEEEpubid{\ieee{}0000--0000/00\$00.00~\copyright~2021 IEEE}

\maketitle

\begin{abstract}	
    This work studies beamformers designed to achieve multiple characteristics simultaneously, specifically those compromising white-noise gain and directivity factor. We compare methods explicitly designed for these joint features against those obtained by combining specific single-task beamformers. Through simulations, we demonstrate that the robust superdirective and the tunable beamformers yield the best results among those studied. Notably, these two methods produced nearly identical outputs across all evaluated metrics. These two are also more practical, continuously compromising between the two objectives.
\end{abstract}

\section{Introduction}

When dealing with fixed beamforming \cite{zhao_application_2011}, the design of beamformers with multiple features is of some interest \cite{ramos_delayandsum_2011,li_robust_2016,madhavanunni_beam_2023}, to attain an acceptable behavior on two (or more) metrics simultaneously. Two of these coworking features are the white-noise gain (WNG) and the directivity factor (DF); the former dictates how well the beamformer deals with incoherent noise present in the sensors, while the later represents how well the beampattern works when within an anisotropic noise field \cite{benesty_microphone_2008,barnov_spatially_2018,cray_directivity_2001}.

Different beamforming techniques can be employed to design a beamformer that compromises these two characteristics, such as the robust superdirective, and the subspace beamformers \cite{mabande_design_2009,benesty_fundamentals_2017}. Both are inherently designed to co-maximize these metrics, giving more importance to one or the other via some parameter. Another possibility is to design two separate beamformers, and then through some synthesis method bring them together to achieve the desired result. Two such synthesis methods are the Kronecker-Product method \cite{yang_design_2019,kuhn_kronecker_2021} and the Convolutive Kronecker-Product method \cite{frank_constantbeamwidth_2022}.

In this work we present the desired techniques within the framework of an uniform linear array (ULA) positioned in an anechoic environment with far-field sources, and through simulations we compare them to find some conclusions regarding their performance and output. Moreover, we would like to compare how well the combination of different beamformers through synthesis methods fares, when put against ones that are purposefully designed with the trade-off objective in mind.
\section{Signal model}

Let $S$ be an uniform linear array (ULA) comprised of $M$ sensors with fixed intersensor spacing $\delta$. This array is placed in a anechoic environment, populated by desired and contaminating sources. In the frequency domain, the observed signal vector is
\begin{equation}
	\bvy(f) = \bvx(f) + \bvv(f),
\end{equation}
where $\bvy(f) = \tr{[y_1(f),\cdots,y_M(f)]}$ are the $M$ observed signals (with similar definition for $\bvx(f)$ and $\bvv(f)$), $\bvx(f)$ is the desired signal, and $\bvv(f)$ is the noise signal. Furthermore, assuming the desired signal's source is in the far-field, and from the anechoic assumption, the desired signal can be written as
\begin{equation}
    \bvx(f) = \bvd{x}(f) x_1(f),
\end{equation}
with $x_1(f)$ being the signal at the reference sensor ($m=1$), and $\bvd{x}(f)$ being the desired signal's steering vector, given by
\begin{subgather}
	\bvd{\t_x}(f) = \tr{\tup{ 1 , \vartheta(f) ,, [\vartheta(f)]^{(M-1)} }}, \\
	\vartheta(f) = e^{ -\j2\pi f\frac{\delta}{c} \cos{\theta_x} },
\end{subgather}
where $\theta_x$ is the desired source's direction w.r.t. the ULA's endfire direction. For simplicity, we assume that $\theta_x = 0$, aligned with the array's axis.

\subsection{Filtering}

The observed signal vector can be used to estimate the desired signal at the reference via a linear filter $\bvh(f)$; that is, the estimate $z(f)$ can be obtained through
\begin{equations}
	z(f) 
	& \approx x_1(f) \\
	& = \he{\bvh}(f) \bvy(f) \\
	& = \he{\bvh}(f)\bvd{x}(f) x_1(f) + \he{\bvh}(f) \bvv(f),
\end{equations}
\IEEEpubidadjcol 
where $\bvh(f)$ is designed to attain some desired beamforming features. Among these, the most traditional ones are the desired signal preservation (achieved through the constraint $\he{\bvh}(f) \bvd{x}(f) = 1$), and the noise rejection (done by minimizing the residual noise $\he{\bvh}(f) \bvv(f)$). Given these two objectives, and assuming a correlation matrix $\Corr{\bvv}(f)$ for the (generic) noise signal, the beamformer $\bvh(f)$ is written as (omitting $(f)$ for conciseness)
\begin{equation}
    \bvh = \argmin_{\bvh} \he{\bvh} \Corr{\bvv} \bvh~\text{s.t.}~\he{\bvh} \bvd{x} = 1,
\end{equation}
whose solution is
\begin{equation}
	\label{eq:sec2:basic_form_beamformer}
	\bvh(f) = \frac{\inv{\Corr{\bvv}}(f) \bvd{x}(f)}{\he{\bvd{x}}(f) \inv{\Corr{\bvv}}(f) \bvd{x}(f) }.
\end{equation}

\subsection{Beamforming metrics}
In this work, we are interested in the trade-off between white-noise gain (WNG), which measures the beamformer's performance under uncorrelated noise; and directivity factor (DF), assessing its capability under an isotropic-noise field. Respectively, these metrics are given by
\begin{subgather}
	\wng[\bvh](f) = \frac{\abs{\he{\bvh}(f) \bvd{x}(f)}^2}{\he{\bvh}(f) \bvh(f)} , \\
	\df[\bvh](f) = \frac{\abs{\he{\bvh}(f) \bvd{x}(f)}^2}{\he{\bvh}(f) \Gzp(f) \bvh(f)} ,
\end{subgather}
in which $\Gzp(f)$ is the isotropic-noise field correlation matrix,
\begin{equation}
	\label{eq:def_Gzp}
	\Gzp(f) = \frac{1}{2}\int_{0}^{\pi} \bvd{\t}(f) \he{\bvd{\t}}(f) \sin{\t} \dd \t.
\end{equation}

Alternatively, these metrics can also be presented in their broadband forms,
\begin{subgather}
	\wng[\bvh] = \pts{\int_{f_0}^{f_1} \frac{1}{\wng[\bvh](f)} \dd f}^{-1} , \\
	\df[\bvh] =\pts{\int_{f_0}^{f_1} \frac{1}{\df[\bvh](f)} \dd f}^{-1}
\end{subgather}
where the beamformer is assumed to be distortionless, and therefore $\he{\bvh}(f) \bvd{x}(f) = 1 ~\forall~ f$.

\section{Trade-off beamforming}

While many techniques are available for designing beamformers, fixed beamforming approaches the problem by assuming the noise field (which dictates $\bvv(f)$) is time-invariable, and has a known constant statistical structure in its correlation matrix. In this section, we present and discuss various literature-available beamformers that compromise between WNG and DF.

\subsection{Basic beamformers}

Given the two target metrics, beamformers can be designed to maximize each of them. Namely, these are the delay-and-sum beamformer, maximizing WNG; and the superdirective beamforming, maximizing DF. Respectively, these are given by \cite{perrot_you_2021,benesty_fundamentals_2017}
\begin{subgather}{eqs:maxwng_and_maxdf_beamformers}
	\Corr{\bvv;\wng}(f) = \Id{M}\label{subeq:maxwng_beamformer} \\
	\Corr{\bvv;\df}(f) = \Gzp(f) \label{subeq:maxdf_beamformer}
\end{subgather}

\subsection{Closed-form beamformers}

From the objective of jointly maximizing WNG and DF, a few closed-form beamformers can be used.

\subsubsection{Robust superdirective}

The first such filter is the robust superdirective beamformer $\bvh{\rsd;\alpha}(f)$ \cite{wang_robust_2023}, which linearly interpolates between the isotropic noise field $\Gzp(f)$ and the uncorrelated white noise field $\Id{M}$ (the identity matrix). Namely, for this beamformer, we let
\begin{equations}{eq:robust_superdirective}
	\Corr{\bvv;\rsd}(f) 
	& \approx \bvGa{0,\pi;\alpha}(f) \\
	& = [1 - \alpha]\Gzp(f) + \alpha \Id{M}
\end{equations}
with $0 \leq \alpha \leq 1$ being the compromising parameter. Comparing to \cref{subeq:maxdf_beamformer}, the addition of $\alpha \Id{M}$ acts as a regularization, ensuring the invertibility of $\Corr{\bvv;\rsd}(f)$, and that a floor-level of white noise rejection is achieved.


\subsubsection{Tunable beamformer}

Another option for a closed-form compromising beamformer is the tunable beamformer $\bvh{\tun;\psi}(f)$ proposed in \cite{berkun_tunable_2016}, where the noise correlation matrix is given by
\begin{subgather}{eq:tunable}
	\Corr{\bvv;\tun}(f) = \bvGa{\psi,\pi} + \epsilon \Id{M} \\
	\epsilon = \frac{1-\cos\psi}{2}
\end{subgather}
where $\bvGa{\psi,\pi}$ is similar to \cref{eq:def_Gzp}, but integrating from $\psi$ to $\pi$; and $\psi$ is a regularization angle, $0 \leq \psi \leq \pi$. Its idea is to ignore narrow-angle directions in the isotropic noise field, along with the addition of regularization.

Given their respective correlation matrices, these beamformers (robust superdirective and tunable) can be directly obtained using \cref{eq:sec2:basic_form_beamformer}. From their controlling parameters ($\alpha$ and $\psi$), it is true that
\begin{subgather}{eqs:sec3:relationship_beamformers_df-wng}
	\bvh{\rsd;0}(f) = \bvh{\sub;1}(f) = \bvh{\tun;0}(f) = \bvh{\df}(f) \\
	\bvh{\rsd;1}(f) = \bvh{\sub;M}(f) = \bvh{\tun;\pi}(f) = \bvh{\wng}(f)
\end{subgather}

Therefore, choosing the parameters' values leads to a compromise between WNG and DF for these beamformers.

\subsection{Combining beamformers}

Along with closed-form solutions, techniques that combine features are also available. Two notable ones are the Kronecker-Product (KP) \cite{yang_design_2019,huang_robust_2020} and Convolutive KP (CKP) \cite{frank_constantbeamwidth_2022} ones. These techniques are based on separating the array into sub-arrays (with specific mechanisms for each technique), designing a beamformer for each sub-array, and then combining these sub-beamformers into a filter for the whole array.

Given our WNG-DF compromise objective, one of the arrays will be a delay-and-sum beamformer, and the other a superdirective one, both presented in \cref{eqs:maxwng_and_maxdf_beamformers}.

\subsubsection{Kronecker-Product beamforming}
\label{subsubsec:kp_beamformer}

We let $S_1$ and $S_2$ be two sub-arrays of $S$, each with $M_1$ and $M_2$ sensors respectively ($M = M_1 M_2$), and steering vectors $\bvd{x,1}(f)$ and $\bvd{x,2}(f)$, such that
\begin{equation}
	\bvd{x}(f) = \bvd{x;1}(f) \otimes \bvd{x;2}(f)
\end{equation}
where $\otimes$ is the Kronecker product operator \cite{loan_ubiquitous_2000}. By designing a beamformer for each sub-array, namely $\bvh{1}(f)$ and $\bvh{2}(f)$, we can obtain the beamformer for the whole ULA as \cite{frank_constantbeamwidth_2022}
\begin{equation}
	\bvh{\lkp;M_1}(f) = \bvh{1}(f) \otimes \bvh{2}(f)
\end{equation}
in which $M_1$ is the trade-off parameter. The sub-filters $\bvh{1}(f)$ and $\bvh{2}(f)$ are respectively obtained via \cref{subeq:maxwng_beamformer,subeq:maxdf_beamformer}. Furthermore, for $\bvh{2}(f)$, we use $\Gzp(f)$ as a $\sz{M_2}{M_2}$ matrix, calculated for the sub-array $S_2$ with \cref{eq:def_Gzp}.

\subsubsection{Convolutive KP beamforming}
\label{subsubsec:ckp_beamformer}

Given again two sub-arrays with $M_1$ and $M_2$ sensors, we now require that $M_1 + M_2 - 1 = M$. In this scenario, $\bvd{x;1}(f)$ is relative to the first $M_1$ elements of $\bvd{x}(f)$, and $\bvd{x;2}(f)$ to the first $M_2$ elements. Thus, we can synthesize $\bvh{\ckp;M_1}(f)$ (the beamformer for $S$) as \cite{frank_constantbeamwidth_2022}
\begin{equation}
	\bvh{\ckp;M_1}(f) = \bvh{1}(f) \ast \bvh{2}(f)
\end{equation}
where $\ast$ symbolizes the (sensor-wise) convolution operation.

The sub-filters $\bvh{1}(f)$ and $\bvh{2}(f)$ are calculated using \cref{subeq:maxwng_beamformer,subeq:maxdf_beamformer} respectively, with $\Gzp(f)$ used for $\bvh{2}(f)$ being calculated with respect to the $S_2$ sub-array. Similarly to \cref{eqs:sec3:relationship_beamformers_df-wng}, here we have the relationships
\begin{subgather}
	\bvh{\lkp;1}(f) = \bvh{\ckp;1}(f) = \bvh{\df}(f) \\
	\bvh{\lkp;M}(f) = \bvh{\ckp;M}(f) = \bvh{\wng}(f)
\end{subgather}

\definecolor{ColA}{HTML}{991F3D}
\definecolor{ColB}{HTML}{997A1F}
\definecolor{ColC}{HTML}{3D991F}
\definecolor{ColD}{HTML}{1F997A}
\definecolor{ColE}{HTML}{1F3D99}
\definecolor{ColF}{HTML}{7A1F99}

{
	\let\fbox\relax
	\begin{figure*}[t]
		\centering\noindent
		\begin{subfigure}{0.32\linewidth}
				\centering

\pgfplotsset{compat=1.18}
\begin{tikzpicture}[tight background,
	]
	\begin{lineplot}{Normalized parameter}{WNG ($\dB$)}[legend to name={lineplot_wng}, ymin=-50, ymax=15, xmin=-0.02, xmax=1.02, xtick={0, 0.5, 1}]
		\addplot [style=styleA, thick]
		table [col sep=comma, y=val] {figures/io_input/wng_rsd.csv};
		
		\addplot [style=styleB, thick]
		table [col sep=comma, y=val] {figures/io_input/wng_sub.csv};
		
		\addplot [style=styleB, thick]
		table [col sep=comma, y=val] {figures/io_input/wng_tun.csv};
		
		\addplot [style=styleC, thick]
		table [col sep=comma, y=val] {figures/io_input/wng_kp.csv};
				
		\addplot [style=styleD, thick]
		table [col sep=comma, y=val] {figures/io_input/wng_ckp.csv};
	\end{lineplot}
\end{tikzpicture}
				\caption{White-noise gain.}
				\label{subfig:lineplot_wng}
		\end{subfigure}
		\fbox{\begin{subfigure}{0.32\linewidth}
				\centering

\usetikzlibrary{backgrounds}
\pgfplotsset{compat=1.18}
\begin{tikzpicture}[tight background,
	]
	\begin{lineplot}{Normalized parameter}{DF ($\dB$)}[legend to name={lineplot_df}, ymin=13, ymax=17, xmin=-0.02, xmax=1.02, xtick={0, 0.5, 1},]
		\addplot [style=styleA, thick]
		table [col sep=comma, y=val] {figures/io_input/df_rsd.csv};
		
		
		\addplot [style=styleB, thick]
		table [col sep=comma, y=val] {figures/io_input/df_tun.csv};
		
		\addplot [style=styleC, thick]
		table [col sep=comma, y=val] {figures/io_input/df_kp.csv};
		
		\addplot [style=styleD, thick]
		table [col sep=comma, y=val] {figures/io_input/df_ckp.csv};
		\addlegendentry{$\rsd$};
		\addlegendentry{$\tun$};
		\addlegendentry{$\lkp$};
		\addlegendentry{$\ckp$};
	\end{lineplot}
\end{tikzpicture}
				\caption{Directivity factor.}
				\label{subfig:lineplot_df}
		\end{subfigure} }
		\fbox{\begin{subfigure}{0.32\linewidth}
				\centering

\pgfplotsset{compat=1.18}
\begin{tikzpicture}[tight background,
	]
	\begin{lineplot}{WNG ($\dB$)}{DF ($\dB$)}[legend to name={lineplot_wxd}, ymin=13, ymax=17, xmin=-50, xmax=15]
		\addplot [style=styleA, thick]
		table [col sep=comma, y=df] {figures/io_input/wxd_rsd.csv};
		
		
		\addplot [style=styleB, thick]
		table [col sep=comma, y=df] {figures/io_input/wxd_tun.csv};
		
		\addplot [style=styleC, thick]
		table [col sep=comma, y=df] {figures/io_input/wxd_kp.csv};
		
		\addplot [style=styleD, thick]
		table [col sep=comma, y=df] {figures/io_input/wxd_ckp.csv};
	\end{lineplot}
\end{tikzpicture}
				\caption{WNG x DF.}
				\label{subfig:lineplot_wxd}
		\end{subfigure}} \\
		\vspace*{1em}
		\ref*{lineplot_df}
		\caption{Results for different parameter values.}
		\label{fig:lineplots_sweep}
	\end{figure*}
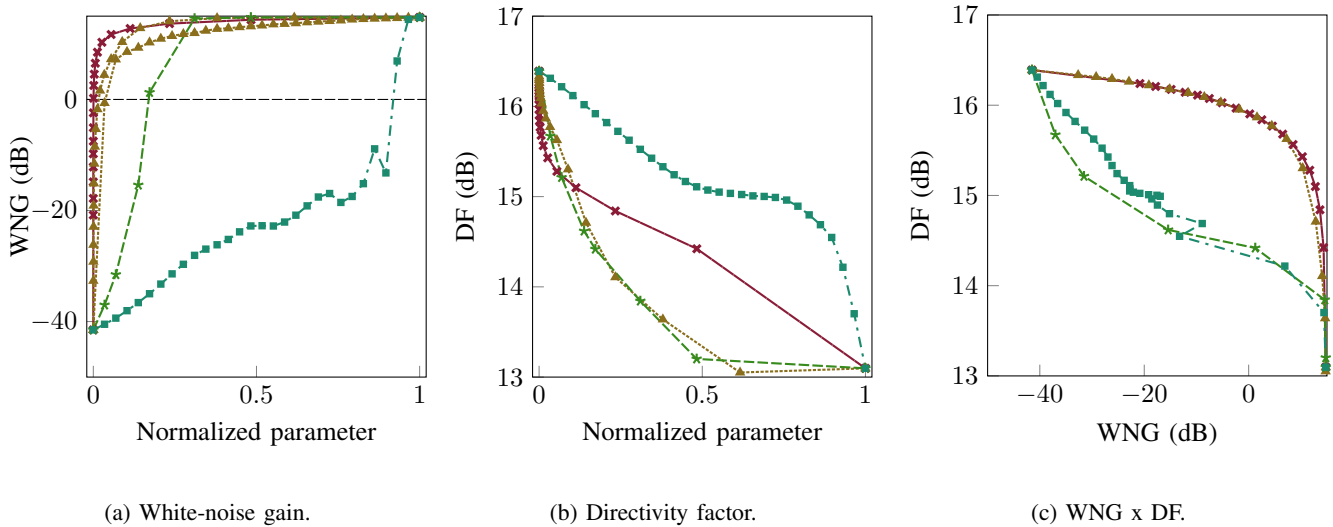
}

\section{Simulations and discussion}

In these simulations, we will compare all beamformers previously presented. We assume that the linear array is comprised of $M = 30$ sensors, with spacing $\delta = 2~\si{\centi\meter}$. This number of sensors was chosen given its number of divisors (fundamental for the $\lkp$ beamformer), and this spacing to ensure the absence of aliasing effects \cite{dmochowski_spatial_2009}, which (in this configuration) happens at $f \approx 8.5\si{\kilo\hertz}$.

Four beamformers will be tested:
\vspace*{-1em}
\begin{itemize}
	\item RSD: A robust superdirective, based on \cref{eq:robust_superdirective};
	\item TUN: A tunable beamformer, using \cref{eq:tunable};
	\item KP: A Kronecker product based one, as detailed in \cref{subsubsec:kp_beamformer};
	\item CKP: A convolutive Kronecker product based beamformer, following \cref{subsubsec:ckp_beamformer}.
\end{itemize}

The results in \cref{fig:lineplots_sweep} show the beamformers' broadband performance on the range $200-8\si{\kilo\hertz}$, in terms of WNG and DF. \Cref{subfig:lineplot_wng,subfig:lineplot_df} show the performance for each parameter value (these being normalized on the range $[0,1]$), while \cref{subfig:lineplot_wxd} presents a WNG vs DF plot.  Given the controlling parameter for the two Kronecker product based beamformers (KP and CKP) is discrete, all possible values for each parameter were tested.

From these results, we first see that all beamformers fulfilled the objective of compromising between the two metrics. From \cref{subfig:lineplot_wxd} it is clear that the RSD and TUN beamformers offer the best compromise between the two metrics (largest WNG for any given DF, and vice-versa), with the RSD being marginally better; the two Kronecker-product beamformers had similar performance. Further exploring the latter two techniques, from \cref{subfig:lineplot_wng,subfig:lineplot_df} we see that while the KP beamformer quickly reached near-maximum WNG, the CKP one had a more steady rise. The CKP beamformer also had a slightly better performance.

We can also compare the beamformers in terms of practical application: while the KP and CKP filters have discrete controlling parameters (this being even more important on the KP beamformer, as it depends on the factors of the number of sensors), the other two (RSD and TUN) have continuously varying parameters, allowing for a precise and flexible selection on the trade-off.

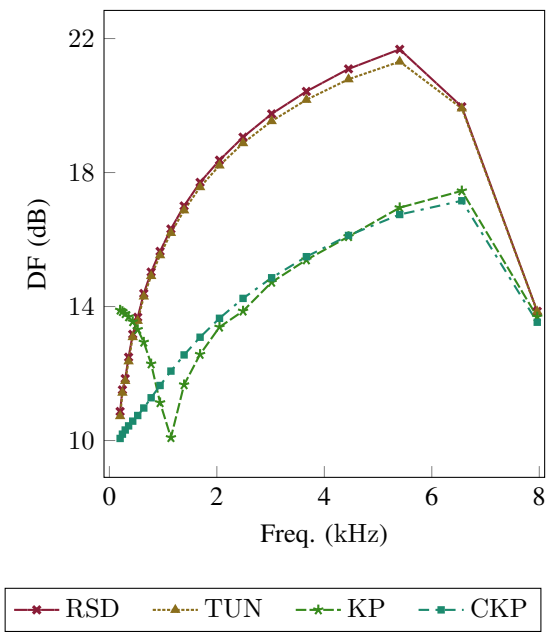
\begin{figure}[t]
	\centering
	\begin{minipage}{0.8\linewidth}
		\centering

\pgfplotsset{compat=1.18}
\begin{tikzpicture}
	\begin{lineplot}{Freq. ($\si{\kilo\hertz}$)}{DF ($\dB$)}[legend to name={lineplot_sameWNG_wng}, xtick={0, 2000, 4000, 6000, 8000}, xticklabels={0, 2, 4, 6, 8}, xmin=-100, xmax=8100, ytick={10, 14, 18, 22}]
		\addplot [style=styleA, thick] 
		table [col sep=comma, y=val] {figures/io_input/freq_df_rsd.csv};

		
		\addplot [style=styleB, thick]
		table [col sep=comma, y=val] {figures/io_input/freq_df_tun.csv};
		
		\addplot [style=styleC, thick]
		table [col sep=comma, y=val] {figures/io_input/freq_df_kp.csv};
		
		\addplot [style=styleD, thick]
		table [col sep=comma, y=val] {figures/io_input/freq_df_ckp.csv};
		\addlegendentry{$\rsd$};
		\addlegendentry{$\tun$};
		\addlegendentry{$\lkp$};
		\addlegendentry{$\ckp$};
	\end{lineplot}
\end{tikzpicture}
		\label{subfig:lineplot_sameWNG_wng} \\
	\end{minipage}
	\vspace*{1em}
	\ref*{lineplot_sameWNG_wng}
	\caption{Results for similar WNG performance.}
	\label{fig:lineplots_sameWNG}
\end{figure}

In \cref{fig:lineplots_sameWNG}, we present the DF given a fixed WNG (at around $1.2\dB$) over the spectrum, in the aforementioned range $200-8\si{\kilo\hertz}$. This cements that the robust superdirective and the tunable beamformers not only have a better broadband performance, as well as a better per-frequency output, given a fixed frequency.

These results align with the concept that, although useful from the perspective of amalgamating different tools to achieve a joint outcome, instruments such as the KP and CKP methods aren't a match for purposefully designed implementations, in terms of achieving the same goals. They may be useful in scenarios were such implementations aren't available, but these fusing mechanisms are outclassed by more intelligently and deliberately constructed beamformers.

\section{Conclusion}

In this work, four different compromising beamformers that trade between white-noise gain and directivity factor were studied, obtained through widely different frameworks. While two were closed-form beamformers, treating the problem via regularization, the other two were obtained through techniques that combine different beamforming designs.

The results show that, while maybe useful in scenarios where a straightforward solution isn't available, beamforming fusing approaches are outperformed by closed-form solutions, both in narrow- and broadband analyses. These combining techniques also lack flexibility, with discrete compromising parameters.

We also showed that, in practice, robust superdirective and tunable beamformers (both closed-form approaches) are practically identical in performance, with the robust superdirective having a minimal edge.
\printbibliography[]
%
%


\end{document}